%% file: camera_ready.tex
\documentclass[conference]{IEEEtran}
\IEEEoverridecommandlockouts
\usepackage{cite}
\usepackage{amsmath,amssymb,amsfonts}
\usepackage{algorithm}
\usepackage{algorithmic}
\usepackage{graphicx}
\usepackage{textcomp}
\usepackage{xcolor}
\usepackage{float}
\usepackage{svg}
\usepackage{soul}
\usepackage{booktabs}
\usepackage{todonotes}
\usepackage[citecolor=blue,colorlinks=true,linkcolor=blue]{hyperref}%
\usepackage{array}

\usepackage{booktabs}
\usepackage{threeparttable}
\usepackage{makecell}  
\usepackage{amsmath}

\def\BibTeX{{\rm B\kern-.05em{\sc i\kern-.025em b}\kern-.08em
    T\kern-.1667em\lower.7ex\hbox{E}\kern-.125emX}}
\begin{document}

\title{LogicSparse: Enabling Engine-Free Unstructured Sparsity for Quantised Deep-learning Accelerators\\
\thanks{This work was supported by the Sustainable Energy Authority of Ireland under Grant number 24/RDD/1170. 
}}
\author{
Changhong Li, 
Biswajit Basu,
Shreejith Shanker \\ 
Reconfigurable Computing Systems Lab, Electronic \& Electrical Engineering\\
Trinity College Dublin, Ireland\\
Email: \{lic9, basub, shankers\}@tcd.ie
}

\maketitle

\input{p_abstract}
\input{p_introduction}

\input{p_methodology}
\input{p_results}

\bibliographystyle{unsrt}
\bibliography{references}

\end{document}

%% file: p_abstract.tex
\begin{abstract}
FPGAs have been shown to be a promising platform for deploying Quantised Neural Networks (QNNs) with high-speed, low-latency, and energy-efficient inference.
However, the complexity of modern deep-learning models limits the performance on resource-constrained edge devices.
While quantisation and pruning alleviate these challenges, unstructured sparsity remains underexploited due to irregular memory access.
This work introduces a framework that embeds unstructured sparsity into dataflow accelerators, eliminating the need for dedicated sparse engines and preserving parallelism.
A hardware-aware pruning strategy is introduced to improve efficiency and design flow further.
On LeNet-5, the framework attains 51.6$\times$ compression and 1.23$\times$ throughput improvement using only 5.12\% of LUTs, effectively exploiting unstructured sparsity for QNN acceleration.
\end{abstract}

%% file: p_introduction.tex
\section{Introduction}\label{introduction}
The growing complexity of deep learning (DL) models poses significant challenges to hardware resources and energy efficiency.
To address these issues, model compression techniques such as pruning and quantisation have been widely adopted. Previous research, such as Deep Compression~\cite{han2015deep}, demonstrated that these techniques can be effectively combined in a complementary manner, achieving significant compression ratios without sacrificing accuracy.
Owing to their inherent hardware-friendliness, quantisation and structured pruning are widely utilised on GPUs/FPGAs. 

Unstructured pruning removes individual weights, offering greater flexibility and a higher theoretical compression ratio upper bound
with minimal or no loss in accuracy.
Nevertheless, unstructured pruning introduces irregular
memory access patterns, posing challenges such as reduced utilisation of compute units and increased memory access overhead.
As a compromise, $N{:}M$ sparsity has become the widely supported format in mainstream hardware platforms such as NVIDIA’s recent GPU architectures and AMD’s Vitis-AI tools.
However, the effective hardware utilisation of unstructured sparsity is still in the exploratory stage.
To leverage the advantages of unstructured sparsity on FPGAs, many recent works have focused on designing sparse matrix multiplication operators tailored for FPGAs~\cite{montgomerie2023pass,munoz2023flexagon,luo2024sparm}.
However, such sparsity engines often involve complex runtime scheduling and additional control logic which may undermine the inherent parallelism of dataflow deep learning architectures.

\begin{figure}[t!]
    \centering
    \includegraphics[width=0.48\textwidth]{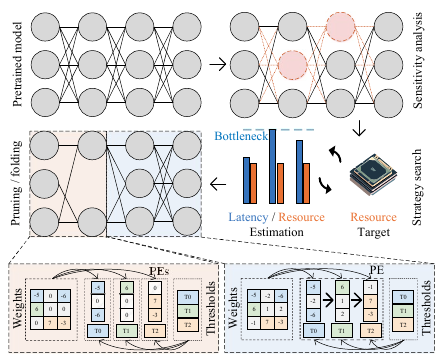}  
    \caption{Workflow of automated pruning and folding decisions}
    \label{fig:framework}
\end{figure}
Recent studies have shown that fine-grained pruning with mixed-precision quantisation can greatly enhance fully unrolled DNNs~\cite{umuroglu2020logicnets, dai2024kratos}. However, existing approaches rely on fixed pruning patterns, limiting flexibility and scalability.
This work proposes an automated pruning workflow for dataflow-based DNN frameworks. 
By statically mapping sparse connections and embedding sparsity into FINN’s configuration heuristics, it preserves pipeline parallelism while enabling fine-grained, hardware-aware pruning with minimal overhead.

    
    


%% file: p_methodology.tex
\section{Methodology and Design} \label{methodology}
Bottleneck layers within the DL model are responsible for the higher latency and lower throughput of dataflow-style QNN accelerators. 
Balancing inter-layer performance can be achieved by optimising folding configurations.
Mapping tools achieve this by changing the layer's Processing elements (PEs) and Single Instruction, Multiple Data (SIMD) width.
While traditional design space exploration (DSE) for these optimisations is throughput-oriented (such as in FINN~\cite{blott2018finn}), our DSE further incorporates resource awareness and sparsity into the design space to enable hardware–software co-design.

As shown in Fig.~\ref{fig:framework}, our DSE first performs global magnitude pruning as a reference.
The heuristic folding search with secondary relaxation is then applied to establish a balanced baseline.
If any layer shows lower resource utilisation after sparse-unfolding, it is directly applied.
Otherwise, bottlenecks are iteratively eliminated.
In each iteration, the layer-wise latency and resource usage are estimated from the ONNX graph.
The latency bottleneck is mitigated by applying sparse or factor unfolding to enhance overall performance under resource constraints.
This process continues until no new optimisation strategy satisfies the overall resource constraint.

Finally, the folding and sparse layer configuration are generated.
The final folding configuration is then adopted for accelerator generation.
Layers that have the potential to be fully unrolled and sparsified are chosen for re-sparse fine-tuning, whereas those that are determined unsuited for exploration are maintained in dense form to preserve accuracy.
This iterative scheme progressively eliminates performance bottlenecks by jointly exploiting folding and sparsity, thereby achieving superior inter-layer balance.
The proposed method thus improves hardware efficiency, enables a hardware–software co-pruning strategy, and advances the design’s Pareto frontier.

%% file: p_results.tex
\section{Experimental Results}\label{results}
\begin{figure}[t!]
    \centering
    \includegraphics[width=0.485\textwidth]{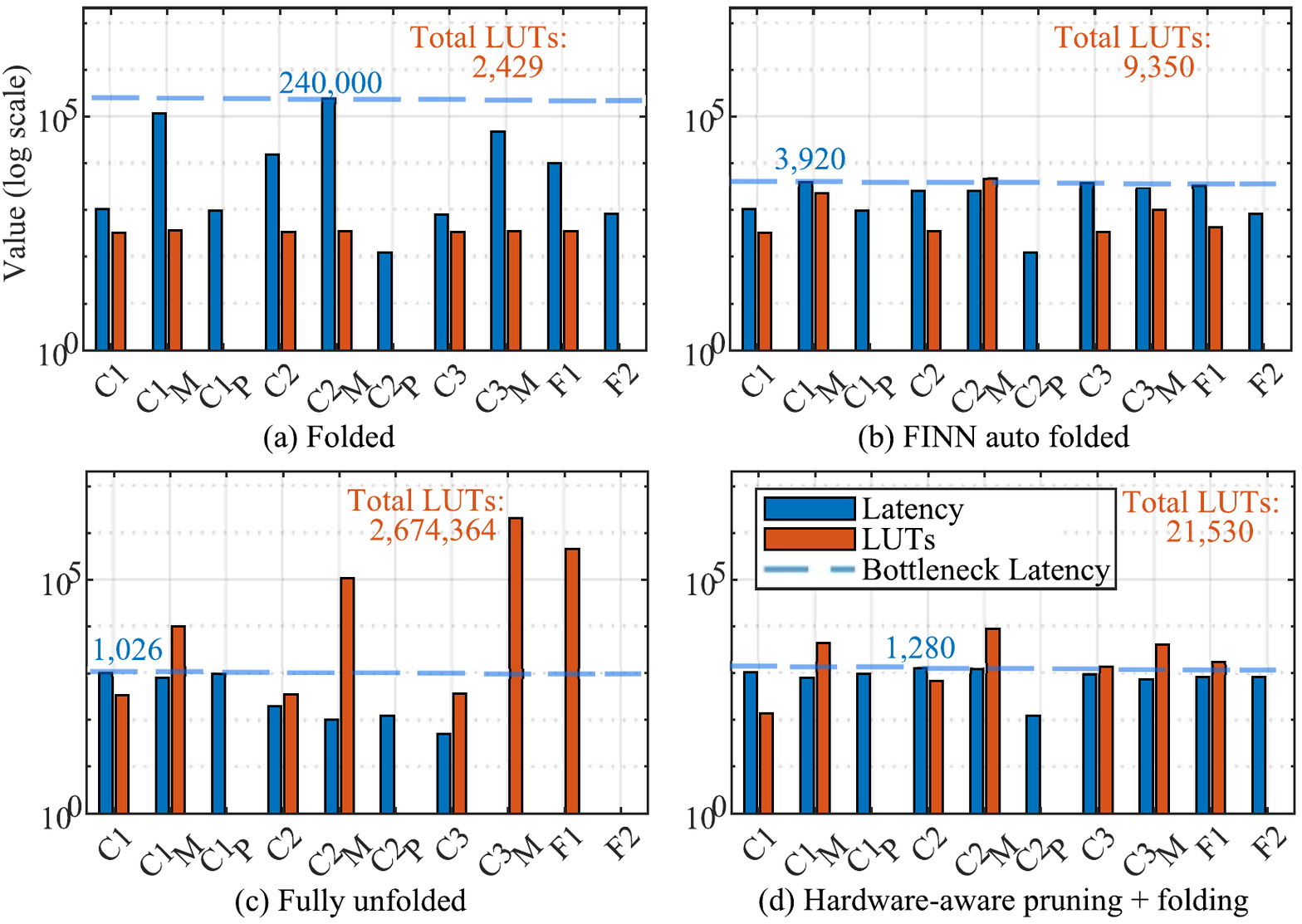} 
    \caption{Estimated latency and LUT utilization per layer of LeNet-5 under different folding and pruning strategies}
    \label{fig:bottleneck} 
\end{figure}

We evaluated our LeNet-5 accelerator on the XCU50 FPGA.
ONNX graphs are used to perform fast latency and resource bottleneck estimation of each layer in different designs, as shown in Fig.~\ref{fig:bottleneck}. For the fully folded network, the second convolutional layer constitutes the major bottleneck. In the automatic unfolding scenario, this bottleneck is significantly alleviated. Fully unrolling the network achieves the lowest bottleneck latency but at the cost of roughly a 1,300$\times$ increase in resource usage.
Under our DSE, the first convolutional layer was further identified as the initial bottleneck and is fully unrolled with unstructured pruning. Subsequently, several fully connected layers, which are identified as the next bottlenecks, are partially unrolled under resource constraints. As a result, our design achieves performance close to the fully unrolled configuration, while consuming significantly fewer resources.

\begin{table}[t!]
\centering
\caption{Performance and resource utilisation comparison of accelerators for LeNet-5 on MNIST}
\label{tab:mnist_lenet5_combined}
\begin{tabular}{lcccc}
\toprule
Work & \makecell{Accuracy\\(\%)} & \makecell{Latency\\($\mu$s)} & \makecell{Throughput\\(FPS)} & \makecell{LUT\\utilisation} \\
\midrule
Rama et al. \cite{yanamala2024empowering} & 98.89  & 1,565   & 995      & 35,644  \\
FPGA-QNN \cite{tasci2025fpga}       & 95.40  & 1,380   & 6,816    & 44,000  \\
Auto folding                         & 98.91  & 44.67   & 65,731   & 9,420   \\
Auto+Pruning                         & 97.78  & 44.56   & 65,866   & 8,553   \\
Unfold                               & 98.91  & 18.18   & 214,919  & 433,249 \\
Unfold+Pruning                             & 97.78  & 15.52   & 251,265  & 100,687 \\
\textbf{Proposed}                    & \textbf{97.82}  & \textbf{18.13}   & \textbf{265,429}  & \textbf{23,465} \\
\bottomrule
\end{tabular}
\end{table}
The measured performance and resource usage of these design strategies are shown in Table~\ref{tab:mnist_lenet5_combined}.
Accelerator designed with proposed DSE, achieves superior latency and throughput compared to a fully unfolded dense accelerator, with only approximately 5\% of the LUT. 
Furthermore, compared with other baseline accelerators, our design achieves the best latency and throughput,  with minimal LUT utilisation, surpassing state-of-the-art implementations. These results demonstrate that our automated framework effectively leverages unstructured sparsity and achieves a well-balanced trade-off between unfolding and sparsity patterns under hardware resource constraints.